# Preference for meat is not innate in dogs

Anandarup Bhadra[1] and Anindita Bhadra[1*]

[1] Behaviour and Ecology Lab, Department of Biological Sciences,
Indian Institute of Science Education and Research – Kolkata, India

[*]**Address for Correspondence:**

Behaviour and Ecology Lab, Department of Biological Sciences,
Indian Institute of Science Education and Research – Kolkata
P.O. BCKV Main Campus, Mohanpur,
Nadia, PIN 741252, West Bengal, INDIA
*tel.* 91-33-25873119
*fax* +91-33-25873020
*e-mail:* abhadra@iiserkol.ac.in

**Abstract**

Indian free-ranging dogs live in a carbohydrate-rich environment as scavengers in and around human settlements. They rarely hunt and consequently do not encounter rich sources of protein. Instead they have adapted to a diet of primarily carbohydrates. As descendents of the exclusively carnivorous wolves, they are subjected to the evolutionary load of a physiological demand for proteins. To meet their protein needs they resort to a thumb rule – if it smells like meat, eat it. Pups face high competition from group and non-group members and are in a phase of rapid growth with high protein demands. Following the thumb rule, then they can acquire more protein at the cost of increased competition and reduced supplementary non-protein nutrition. However, if the mother supplements their diet with protein-rich regurgitates and/or milk, then the pups can benefit by being generalists. Using a choice test in the field we show that while adults have a clear preference for meat, pups have no such preference, and they even eat degraded protein eagerly. Thus the thumb rule used by adult dogs for efficient scavenging is not innate, and needs to be learned. The thumb rule might be acquired by cultural transmission, through exposure to meat in the mother's regurgitate, or while accompanying her on foraging trips.

## Introduction

Adult food preferences in mammals are shaped by genetic predispositions (Scott 1946; Nachman 1959) and by subsequent learning experiences (LeMagnen 1967; Rozin 1967). For example, the flavour of mother's milk provides cues such that the pups preferentially eat what the mother did in rats (Galef and Henderson 1972) and also in pigs (Campbell 1976). The swallowing of amniotic fluid before birth seems to affect food preference in the adult stage in humans (Mennella and Beauchamp 1994) and in sheep (Mistretta and Bradley 1983), suggesting that learning can begin even before birth. The peripheral gustatory system of puppies is already developed at birth but does not reach the adult form until later in life (Ferrell 1984a), such that genetic predispositions can constrain taste perception. Early experiences of food also seem to have an impact on dog food preference (Kuo 1967; Mugford 1977; Ferrell 1984b) which is strongly influenced by the mother, through offering regurgitated partly digested food before weaning (Thorne 1995) and also through foraging in the presence of the pup. Besides the possibility of the strong influence of mother's diet on pups, the pup's own experience also shapes its diet. Evidence of learning has been seen in dogs where flavor experience and physiological effect are well separated in time such that classical conditioning is inadequate for an explanation (McFarland 1978). Neophobia or fear of something new is uncommon in dogs, but it has been reported in case of food (Thorne 1995). Neophilia or preference for something new, is common when it comes to food (Mugford 1977; Griffin et al. 1984). Aversion develops rapidly for food which have a negative physiological response, as has been demonstrated in coyotes (Ellins et al. 1977) and to a lesser degree in dogs (Rathore 1984). So a pup's food preferences may be innate, conditioned by experience or learned either through cultural transmission from the mother or through active teaching by her.

Wolves hunt for meat and occasionally scavenge (Mech and Boitani 2003; Forbes and Theberge 1992), while their modern-day descendents – the pet dogs are fed by their owners in controlled amounts, often leading to over feeding (German 2006; Edney and Smith 1986; McGreevy and Thomson 2005). Free-ranging dogs exist in many countries like Mexico (Ortega-Pacheco et al. 2007; Daniels and Bekoff 1989), Ecuador (Kruuk and Snell 1981), Zambia (Balogh 1993), Zimbabwe (Butler et al. 2004), Italy (Boitani 1983; Bonanni et al. 2010), India (Pal 2001; Vanak and Gompper 2009), Nepal and Japan (Kato and Yamamoto 2003) etc. While they do occasionally hunt and beg for food, they principally acquire food by scavenging (Vanak and Gompper 2009; Vanak et al. 2009; Spotte 2012), making them an ideal model system to study the effects of the earliest form of domestication. Indian free-ranging dogs have appeared in many ancient Indian texts and folklore over the ages, sometimes as a domesticated animal and sometimes as a stray (Debroy 2008). They have lived in their current state in India for generations and are thus well adapted to the scavenging lifestyle that they lead as an integral part of the human ecology today (Pal 2001). Indian free-ranging dogs don't often encounter meat during scavenging in waste dumps and while begging for food. Instead, they live on a carbohydrate-rich omnivorous diet consisting of biscuits, breads, rice, lentil, fish bones, occasional pieces of decomposing meat from a carcass (and even mangoes, cow dung and plastic; personal observations). These dogs have adapted to their scavenging habit without actually giving up the preference for meat (Houpt et al. 1978; Bhadra et al, unpublished data). A possible mechanism might have been the development of better digestion of carbohydrates which has now been demonstrated to be one of the major genetic changes that the ancestors of dogs underwent during their transition from wolves (Axelsson et al. 2013). Given the carbohydrate-rich diet of these dogs, this would be an advantage in terms of meeting their energy requirements, especially in areas like India where the human diet is chiefly comprised of carbohydrates (Mohan et al. 2009). However, it seems that the dogs have behaviourally adapted to scavenging in and around human habitation by developing a thumb rule for foraging - “if it smells like meat, eat it”. This would enable them to always choose the food with a higher intensity of meat smell first, thus helping them sequester higher amounts of protein in their diet (Bhadra et al, unpublished data). We wanted to test the hypothesis that this thumb rule is an innate characteristic of the dogs, and does not need to be learned.

## Materials and Methods

We used the One Time Multi-option Choice Test (OTMCT) module for our experiment (Bhadra et al, unpublished data). A random dog was provided with three food options simultaneously such that all three were equally accessible. All events including the inspection and eating of the food options were recorded in order. The data for only those cases where all the options were at least inspected were used for analysis. These dogs, living in a highly competitive environment, could be expected to eat the preferred food first, and so we recorded the order in which the food was consumed. The experiments were conducted in Kolkata (22°34’10.92” N, 88°22’10.92” E) West Bengal, India, between December 2011and March 2012. In the OTMCT experiments the quantity of food was too small (less than 10 ml) to be a stimulus – we used small lumps of food, approximately the size of an almond. The options were provided such that they were visually identical and the only cue for the dogs to make the choice was the odour of each option. Each dog was given the choice test only once to eliminate the effect of learning and get a clear representation of the preference already formed at the population level. The experiment was conducted in two sets, one with adult dogs and the other with pups aged between 8 to 10 weeks. This age window was chosen because the pups learn to take solid food from external sources, begin exploring by themselves and wean at this age (Pal 2008). In each set, our final sample size was 60.In the experimental set (Experiment 1A), the pups were given a gradient of proteins in novel food. The options provided in OTMCT were P1 (Dog Biscuit - 80% Protein); P2 (Fresh Pedigree - 24%); P3 (One day old Pedigree - Protein degraded) (Please see OSM for details). In the control set (Experiment 1B), adults were given the same choice test. The dog biscuit actually contained some meat and pedigree only contains synthetic protein. The dogs often have to search for food amidst rotting garbage, so it is important for them to discern between fresh and degraded protein. We used the stale pedigree as a source of degraded protein. Neither the pups nor the adults are likely to have been exposed to pedigree or dog biscuit. The adults are known to discern between food options by smell (Houpt et al. 1978) and should thus treat the options differently. Since adults follow the thumb rule, they should prefer the dog biscuit with the meat smell and avoid the stale protein. So for adults, we expected the order of preference to be P1 > P2 > P3. We hypothesized that the juveniles should follow the same order of preference as the adults if the thumb rule is innate.

Absolute choice was defined as the total number of times each option was chosen in a particular experiment. Choice was taken as the complete consumption of a particular option. Eating order was computed for each experiment. A 3x3 matrix was constructed with the three options in the columns and the number of times each option was chosen first, second and third respectively in the rows. Now, a contingency chi-squared test was carried out to determine whether the tables were random. If they were significantly different from random, then the option that was chosen first the highest number of times was taken to be the first preference at the population level. Similarly the options chosen second and third were also determined.

We computed the average ranks for each event in an experiment, thus getting an idea of the order of occurrence of the inspection and eating of each type of food. Each event was assigned a rank based on the order of occurrence. Since there must be 3 inspections in each experiment and 3 possible acts of consumption, each event could receive a rank between 1 and 6. When an event did not occur (one of the options was not consumed) it was assigned the rank of 7, meaning it had a higher rank than if it had been eaten last. The average of all the ranks for each event was calculated.

## Results

From absolute choice, the adults clearly prefer P1 over P2 and P2 over P3 (Two-tailed Fisher's Exact Test; P1-P2: $p = 0.000$, P2-P3: $p = 0.048$ and P1-P3: $p = 0.000$) (Figure 1) whereas the pups prefer all three equally (Two-tailed Fisher's Exact Test; P1-P2: $p = 0.679$, P2-P3: $p = 0.999$ and P1-P3: $p = 0.999$) (Figure 1). In terms of eating order, adults eat P1 first, P2 second and P3 third (chi square = 74.233, df = 4, $p = 0.000$) (Figure 2) (Table 1), while pups eat the food in random order (chi square = 3.797, df = 4, $p = 0.434$) (Figure 2) (Table 1). So pups do not discriminate between different foods (ie, they show neither preference nor aversion) while adults do prefer the meat smell and avoid the food containing degraded protein. The overall rejection rate in adults (96/180) is significantly higher than that in pups (7/180) (Two-tailed Fisher's Exact Test: $p = 0.000$). Hence we reject our null hypothesis, and conclude that the thumb rule is not innate.

This result was corroborated by the average ranks of the eating events, where adults clearly showed a hierarchical order of ranks ($Rank^{P1E} = 4.20 \pm 1.60$, $Rank^{P2E} = 5.85 \pm 1.64$, $Rank^{P3E} = 6.73 \pm 0.63$) (Table 2) and pups did not ($Rank^{P1E} = 4.15 \pm 1.53$, $Rank^{P2E} = 4.37 \pm 1.78$, $Rank^{P3E} = 4.20 \pm 1.75$) (Table 2). All inspections occurred at random order (Experiment 1A: $Rank^{P1I} = 2.93 \pm 1.49$, $Rank^{P2I} = 2.85 \pm 1.69$, $Rank^{P3I} = 2.77 \pm 1.58$; Experiment 1B: $Rank^{P1I} = 2.13 \pm 0.98$, $Rank^{P2I} = 2.27 \pm 1.01$, $Rank^{P3I} = 2.35 \pm 1.36$) (Table 2), but eating only occurred after all the choices had been inspected by the adults (mean of ranks of all inspection for adults is $2.25 \pm 1.13$ and mean of rank of all eatings for adults is $5.59 \pm 1.73$; Two-tailed Mann-Whitney Test: U = 29968.000, DF1 = 180, DF2 = 180, $p = 0.000$). Interestingly, in case of the pups, eating did not begin after all three options had been inspected. The pups seemed to inspect a food item and consume it immediately, before inspecting the next available option. The difference in the average ranks for each pair of inspection and eating was nearly equal to 1 (P1: 1.22 ± 0.74; P2: 1.52 ± 1.33; P3: 1.43 ± 1.05) in case of the pups, while it was more variable (P1: 2.07 ± 1.47; P2: 3.58 ± 1.79; P3: 4.38 ± 1.58) in case of the adults. So we checked how often inspection of a particular food is followed immediately by its consumption, representing a situation when the pups would be driven by their high hunger levels to eat what is edible immediately, without exploring all available options. We called this possible strategy **Sniff and Snatch (SNS)** – this included the cases where the difference between the ranks for eating and inspection of a particular option was 1. 89% of all choices made by pups were a result of this SNS strategy, which was significantly higher than the 63% of the adults (Two-tailed Fisher's Exact Test: $p = 0.000$) (Figure 1).

## Discussion

Adult free-ranging dogs use the thumb rule – if it smells like meat, eat it, for efficient uptake of proteins through scavenging. This thumb rule could be an innate characteristic of the dogs, stabilized through a long history of domestication from wilder ancestors. It is also possible that dogs are not born with the ability to pick out richer sources of protein by smelling meat, but they learn this over time through a process of cultural transmission or by operant conditioning. Our results clearly show that pups (in the weaning stage) do not follow the thumb rule to make a choice of food. On the contrary, they seem to often inspect a particular food followed immediately by its consumption, a strategy which we call sniff and snatch (SNS). Thus we conclude that the thumb rule used by adult dogs is not innate, and needs to be learned at some stage in life.

Our results do not necessarily suggest that the pups are physiologically incapable of distinguishing between food types by the smell. The lack of the thumb rule could also mean that even though pups are physiologically capable of using the thumb rule, they do not, as the requirement for nutrition is too high. The pups are in a phase of rapid growth, have high nutritional needs and may not be able to afford to discriminate between foods simply to meet their calorific demands. Besides, the correlation between resources and number of individuals break down after the breeding season, indicating that the presence of juveniles increases competition within the group (Sen Majumder et al, unpublished data). Parent-offspring as well as sib-sib competition over food given by humans exist, and this typically increases around the time of weaning, when the pups are 8 to 10 weeks old (Paul et al, unpublished data). Thus not only is the intra-group competition high at the level of the population because of resource constraints, but within the group too, the pup has to compete with the mother as well as its siblings because of dynamics of the changing interests of individual group members.

Even in the face of competition, the requirement for specific macronutrients should not decrease in the case of pups that indeed are growing. If anything, given the fact that cellular growth and division is driven by protein production, the requirement for protein in the diet should be higher. Hence even as pups, they can be expected to follow a thumb rule by which they can sequester proteins. The only way in which the thumb rule can become redundant in pups is if their diet is supplemented by proteins from a source which they do not have to find and eat. Pups are fed milk rich in protein by the mother (of the 22.7% dry matter in milk, 9.47% is fat, 7.53% is protein and 3.81% is sugar)(Oftedal 1984). So for the pups, protein:fat:carbohydrate ratio works out to roughly 36%: 45%: 19% compared to the macronutrient content of 30%:63%:7% measured in the diet consumed by adults provided with ad libitum food (Hewson-Hughes et al. 2013). The protein level in milk seems to exceed that required in adults whereas fats (a rich source of energy) seem to be in deficit. As the weaning period approaches, the mother reduces feeding and yet the pups solicit more food. Around 5 to 6 weeks the mother begins to regurgitate solid food (Malm and Jensen 1993; Malm and Jensen 2010) which is also rich in proteins (for a single group, out of 10 observations, 8 contained meat – personal observations, Manabi Paul). This regurgitation from the mother initiates the training for eating solid food. Around 8 to 10 weeks of age the pups begin their own explorations and find food for themselves, but their diet is still supplemented by occasional suckling and regurgitation. Hence they might not need to specifically sequester proteins even if they are capable of doing so since the overall requirement for nutrition is so high. This is substantiated by the fact that the pups reject food at a much lower rate than the adults. The idea is also corroborated by Ontko (1957) who report that "under ad libitum feeding, increased increments of dietary fat in the ration of the weanling dog increased the present protein requirement as measured by rate of growth and by food efficiency". Since the mother's milk provides all the protein required by pups, the nutrient constraints only exist for adults. Under such a situation, predictions of optimal diet theory suggest that partial preferences should develop when fitness is maximized through the rate of food gain maximization subject to some nutrient being maintained at a minimum threshold value (Pyke 1984), as is seen in our experiment. It follows, that pups, not operating under these nutrient constraints, do not need to have this partial preference.

The advantage of not following the thumb rule from the onset of life is manifolds for the dogs. Since resources are patchy, they are in high demand and require defending (Pal et al. 1998). The pups are not capable of such defense and rely on the mother for it. The mother in turn gathers the food, processes and provides it as milk. But she cannot continue this for an indefinite amount of time. To prepare her body for the next breeding cycle she must cut this high cost from her own nutrition budget (Trivers 1974). As a result she gradually switches to regurgitating food which is less costly and occurs less often than providing milk. Finally she stops supplementing the pups' nutrition completely. At this point, the pups have already been exposed to solid food in the form of regurgitated material and have begun exploring and foraging by themselves (Pal 2005). Now the thumb rule can take over. In fact, this late development offers some plasticity to the behavior. The pups may get trained to the most concentrated source of protein in the vicinity as both their mother's regurgitation and their own explorations should expose them to this source. This would then be similar to the observation made in rats (Galef and Henderson 1972) and pigs (Campbell 1976) that develop food preferences based on exposure to certain tastes through their mother’s milk. Thus, the development of a preference for meat in dogs can occur by operant conditioning or cultural transmission, or a combination of the two.

If pups are able to learn the thumb rule without the influence of adults, simply by their own explorations during foraging, operant conditioning would be the more likely mechanism for such learning. However, since pups are typically exposed to regurgitations of the mother and also begin their explorations with her, cultural transmission might play an important role in the learning of the thumb rule, either actively through teaching, or passively through social learning. We intend to carry out controlled experiments to test the importance of these two mechanisms in the development of foraging habits of the dogs.

This study was carried out on random dogs at the level of the population and hence the findings represent a basic fact for all dogs that need to learn to acquire food beyond a point until which the mother is the source of all nutrition. All dogs are subject to the evolutionary load of their ancestors being complete carnivores, hence they have a high requirement for protein (Case et al. 2010). As adults they acquire proteins by hunting, scavenging or begging and have to retain a preference for meat. But as pups, their mother provides the necessary proteins through suckling or regurgitation, and the pups can afford to be generalists in their own foraging bouts. Such a generalist strategy would also serve to minimize competition among siblings over preferred food while foraging as a group, and would benefit the pups by maximizing their calory intake. Hence beginning to forage as a generalist and then learning the thumb rule for specifically sequestering more proteins in their diet should be an evolutionarily stable strategy for the dogs.

Our results not only show that the thumb rule is not innate, but also highlights the importance and influence of early exposure to food for dogs. As their early encounters shape their adult preferences, it emphasizes the role of pet owners in bringing up their dogs. Given that most pups are reared in human homes away from their mothers from a very early age, the diet offered by owners to freshly weaned pups might be crucial in determining the lifetime eating habits of their pets. .

**Acknowledgements**

The experiments were designed and carried out by Anandarup Bhadra (ArB). AB supervised the work and co-wrote the paper with ArB. Tithi Roy helped with the field work during these experiments. Our experiments comply with the regulations for animal care in India. We thanks IISER-K for funding this research.

| Expt. No. | Chi-square value | P value for chi square | Log-likelihood value | P value for log-likelihood | Degrees of Freedom | Option chosen first (no. of times) | Option chosen second (no. of times) | Option chosen third (no. of times) |
|---|---|---|---|---|---|---|---|---|
| 1A (pups) | 3.797 | 0.434 | 3.786 | 0.436 | 4 | - | - | - |
| 1B (adults) | 74.233 | 0.000 | 56.476 | 0.000 | 4 | P1(40) | P2(13) | P3(10) |

**Table 1: The results of the chi square tests performed to check for preference towards different food types provided in Experiment 1A and 1B.**

| Expt. No. | Average Rank (mean ± s.d.) | | | | | |
|---|---|---|---|---|---|---|
| | inspection | | | eating | | |
| | P1 | P2 | P3 | P1 | P2 | P3 |
| 1A (pups) | 2.93 ± 1.49 | 2.85 ± 1.69 | 2.77 ± 1.58 | 4.15 ± 1.53 | 4.37 ± 1.78 | 4.20 ± 1.75 |
| 1B (adults) | 2.13 ± 0.98 | 2.27 ± 1.01 | 2.35 ± 1.36 | 4.20 ± 1.60 | 5.85 ± 1.64 | 6.73 ± 0.63 |

**Table 2: Average rank (mean ± s.d.) for each event (inspection and eating of P1, P2 and P3) in Experiment 1A and 1B.**

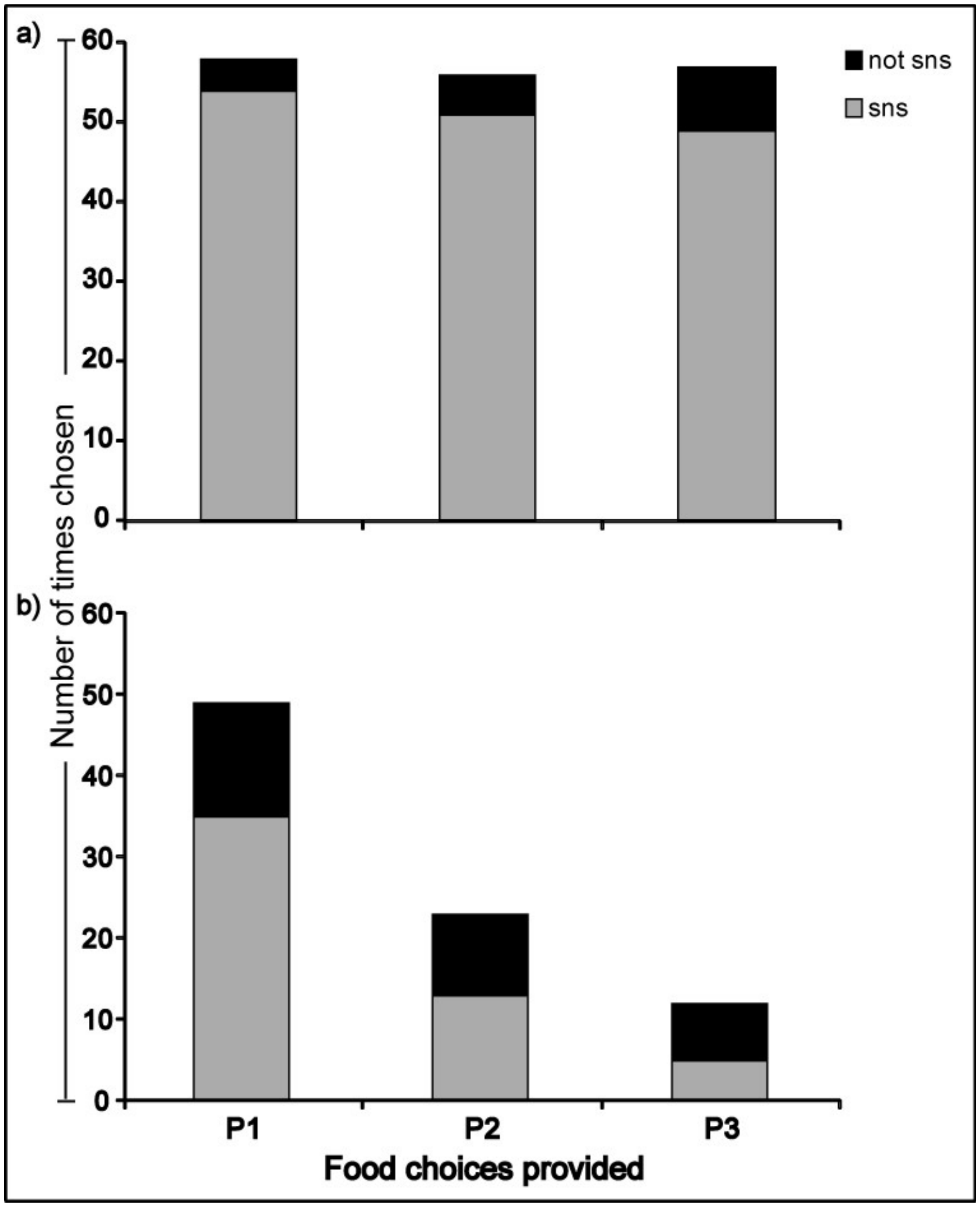

**Figure 1:** Absolute choice data and proportion of SNS from experiment 1A (pups) and B (adults). Absolute choice in experiment 1A, P1 (58) = P2 (56) = P3 (57) (Two-tailed Fisher's Exact Test; P1-P2: $p = 0.679$, P2-P3: $p = 0.999$ and P1-P3: $p = 0.999$) and in experiment 1B, P1 (49) > P2 (23) > P3 (12) (Two-tailed Fisher's Exact Test; P1-P2: $p = 0.000$, P2-P3: $p = 0.048$ and P1-P3: $p = 0.000$). Significant differences are depicted using different alphabets. Proportion of SNS in experiment 1A (154/173) is significantly higher than that in experiment 1B (53/84) (Two-tailed Fisher's Exact Test: $p = 0.000$)

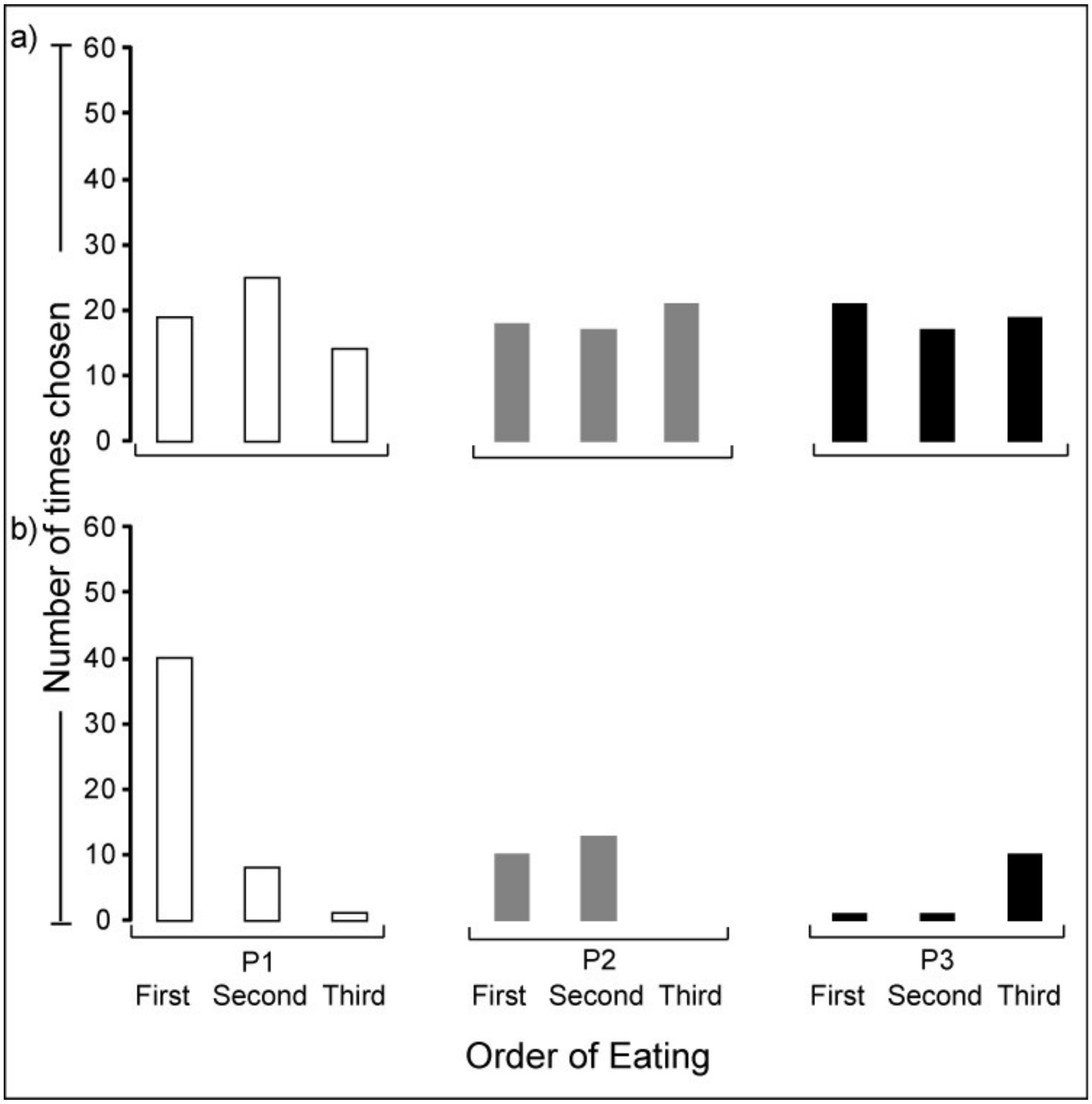

**Figure 2:** Frequency distribution of food choice for determination of eating order in the OTMCT. (A) Pups: Eating order is P1 = P2 = P3 (chi square = 3.797, df = 4, p = 0.434); (B) Adults: Eating order is P1 > P2 > P3 (chi square = 74.233, df = 4, p = 0.000)

**Preference for meat is not innate in dogs**

Anandarup Bhadra and Anindita Bhadra

**Supplementary material**

**Preparation of food options**

We provided three options: P1 (Dog Biscuit - 80% Protein); P2 (Fresh Pedigree - 24%); P3 (One day old Pedigree - Protein degraded). The dog biscuit we used was from "CASTOR & PULLOX PETS SHOP" and nutrition composition was provided in the label (Figure 1). We also used "PUPPY CHICKEN & MILK" from PEDIGREE® for the other two options and again the nutrition composition was provided in the label (Figure 2). None of the options were used after four hours of completion of preparation.

P1: The dog biscuit was soaked in water for an hour, until it became soft and it was then mashed into a paste, such that it became visually identical to P2 and P3.

P2: The pedigree was soaked in water for an hour, until it became soft and it was mashed into a paste, such that it became visually identical to P1 and P2.

P3: The pedigree was soaked in water for at least 24 hours. It was then made into a paste, such that it became visually identical to P1 and P2.

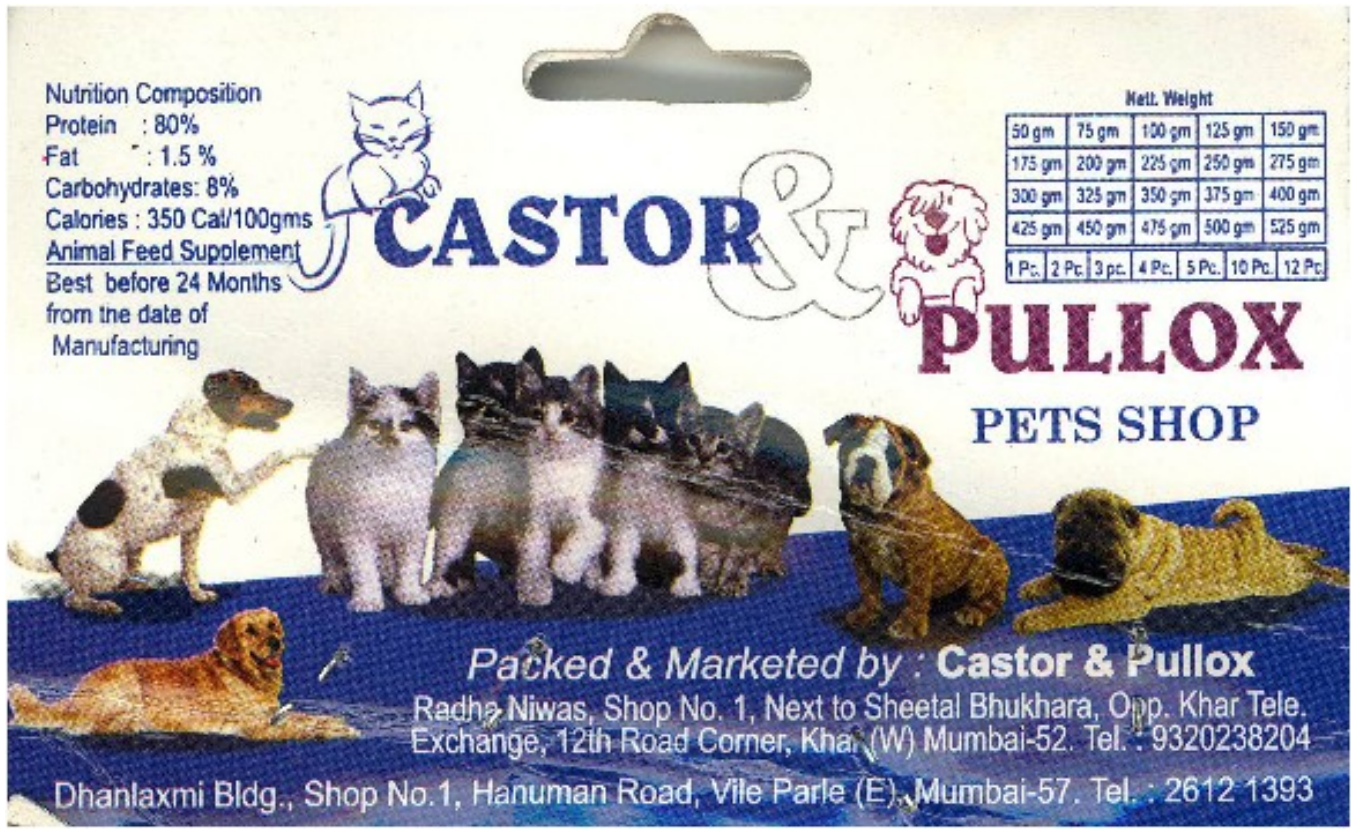

Figure 1

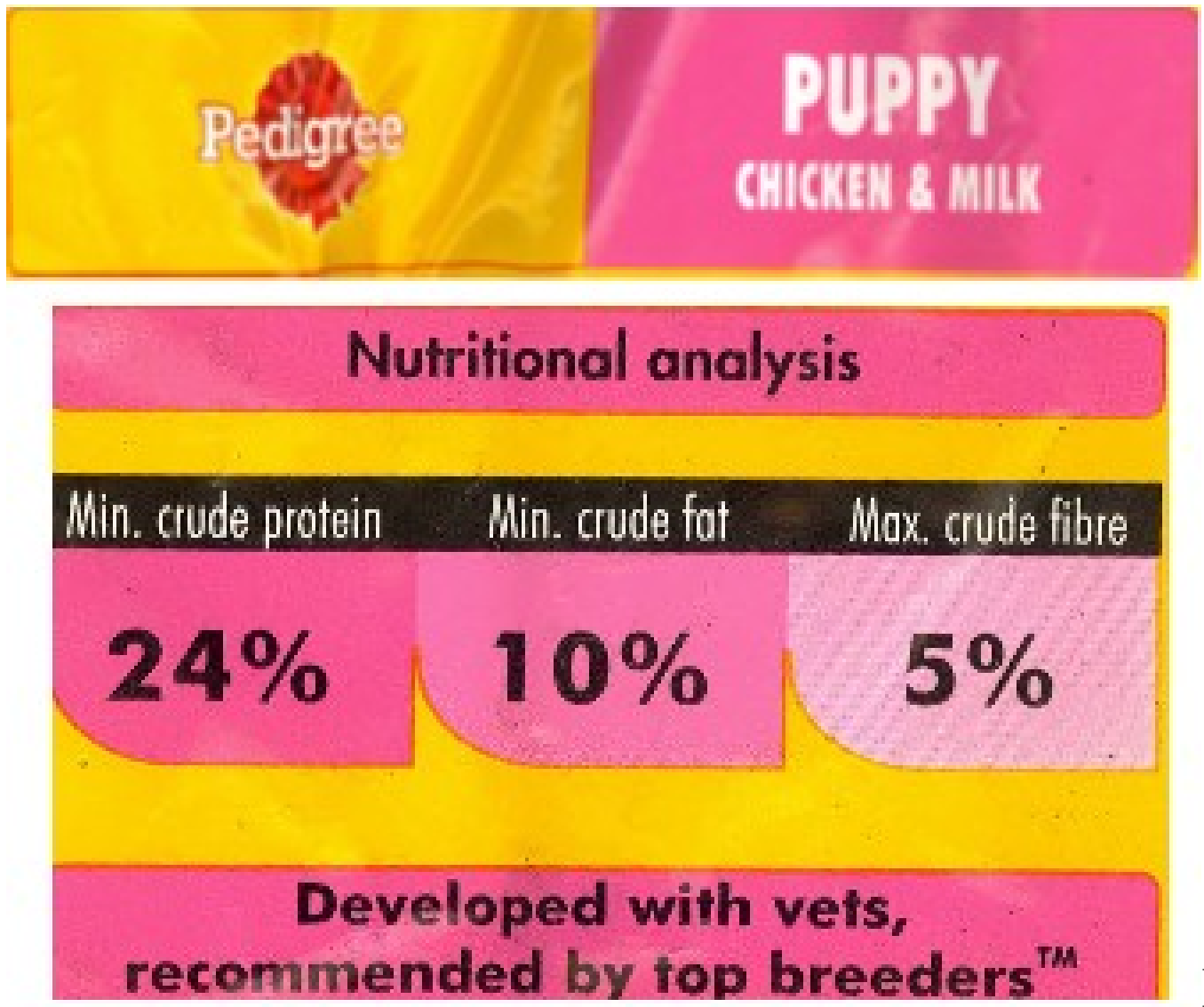

Figure 2